\documentstyle[twocolumn,aps,prl]{revtex}

\begin{document}
\title{{\bf Origin of the photoemission final-state effects in Bi$_{2}$Sr$%
_{2}$CaCu$_{2}$O$_{8}$}\\
{\bf by very-low-energy electron diffraction}}
\author{V. N. Strocov, R. Claessen}
\address{Experimentalphysik II, Universit\"{a}t Augsburg, D-86135 Augsburg,
Germany}
\author{P. Blaha}
\address{Institut f\"{u}r Materialchemie, Technische Universit\"{a}t Wien,
A-1060 Wien, Austria}
\date{\today }
\maketitle

\begin{abstract}
Very-low-energy electron diffraction with a support of full-potential band
calculations is used to achieve the energy positions, ${\bf K}_{\Vert }$
dispersions, lifetimes and Fourier compositions of the photoemission final
states in Bi$_{2}$Sr$_{2}$CaCu$_{2}$O$_{8}$ at low excitation energies.
Highly structured final states explain the dramatic matrix element effects
in photoemission. Intense $c$(2x2) diffraction reveals a significant
extrinsic contribution to the shadow Fermi surface. The final-state
diffraction effects can be utilized to tune the photoemission experiment on
specific valence states or Fermi surface replicas.
\end{abstract}


Photoemission (PE) spectra of Bi$_{2}$Sr$_{2}$CaCu$_{2}$O$_{8}$ (Bi2212) and
other high-temperature superconductors can significantly deviate from the
photohole spectral function $A({\bf k},\omega )$ due to an energy and ${\bf k%
}$ dependence of the matrix element (ME) \cite%
{Lindroos99,Lindroos02,Kordyuk02,Damascelli02}. For example, the long
debated dependence of the shape of the observed Fermi surface (FS) of Bi2212
on the photon energy $h\nu $ and experimental geometry has finally found its
explanation in ME effects. A predictive use of these effects can help to
resolve subtle details of the electronic structure of Bi2212 such as the
bilayer splitting \cite{Lindroos02,Kordyuk02}. Presently, analysis of ME
effects employs computational simulations of the PE process \cite%
{Lindroos99,Lindroos02}. However, there remains an inherent uncertainty of a
few eV due to the lack of any reliable knowledge about the energy location
and lifetime of the PE final states. Additional complications are
incommensurate (IC) structural modulations and antiferromagnetic (AFM)
fluctuations, which manifest themselves in PE experiments, respectively, as
diffraction replicas of the FS and so-called shadow FSs \cite%
{Aebi94,Ding96,Borisenko00}.

Very-Low-Energy Electron Diffraction (VLEED) gives direct access to the PE
final states (see \cite{TMDCs,CFS} and the references therein) based on the
fact that, neglecting the electron-hole interaction within the sudden
approximation, the PE final states are the time-reversed LEED states \cite%
{Feibelman74}. In the VLEED elastic electron reflectivity spectra $R(E)$,
the energies of the spectral structures give the final-state dispersions $E(%
{\bf k})$, and their broadening the corresponding lifetimes $\tau $
expressed by the imaginary part $%
\mathop{\rm Im}%
\Sigma $ of the self-energy as $\hbar /\tau =2%
\mathop{\rm Im}%
\Sigma $.

We here use VLEED to achieve first direct information on the PE final states
in Bi2212 up to 45 eV above the Fermi level $E_{F}$ and elucidate its
immediate implications for the PE experiment.

{\em Experiment and band calculations} -- The VLEED experiment was carried
out using a standard LEED setup operated in the retarding field mode \cite%
{CFS,Krasovskii02}. The total $R(E)$ integrated over all diffracted beams
(plus structureless inelastic reflectivity) was measured in the target
current $T(E)$. Intensities of individual beams were measured on the LEED
screen using a CCD camera. Three samples with an oxygen content near optimal
doping showed identical results.

The band calculations employed the full-potential APW+local orbitals method
as implemented in the WIEN2k package \cite{Blaha01}. The generalized
gradient approximation (GGA) was used to describe exchange-correlation
within the density functional theory (DFT). The calculations employed the
body-centered-tetragonal (bct) crystal structure and a basis of 2200
augmented plane waves. The obtained $E({\bf k})$, characterizing essentially
the bulk states with infinite lifetime, was connected to the VLEED\ process
based on Fourier expansion of the Bloch waves, which was used to evaluate
the partial currents $T_{{\bf k}}$ into the crystal excited in each Bloch
wave, and the integral $T(E)$ spectra \cite{CFC-Vg}. Finite electron
lifetime was simulated by Lorentzian smoothing and damping of $T(E)$ \cite%
{TMDCs}.

The experimental structural parameters from \cite{Tarascon88} and \cite%
{Sunshine88} were tested in the calculations, with the latters yielding the
best agreement with the VLEED experiment. While the valence bands changed
insignificantly, the unoccupied $E({\bf k})$ was strongly sensitive to the
structural parameters. Such a sensitivity suggests that any model of the
final states, used in PE simulations, needs to be checked against the VLEED
experiment.

{\em Integral }$T(E)${\em \ spectra: Dispersions and lifetimes of the final
states} -- The structures in $T(E)$ all reflect those in elastic $R(E)$,
with the inelastic reflectivity giving only a featureless background. We
start with analysis of the normal incidence spectrum in Fig.1({\it a}). It
corresponds, by parallel momentum conservation and assuming insignificant
umklapp contributions from the IC superstructure, to the $\Gamma Z$
direction of the Brillouin zone (BZ). A slope of $T(E)$ towards higher
energies is due to an increase of the inelastic reflection. Its unusual
steepness may indicate strong electron coupling in the cascade relaxation
process. To emphasize the $T(E)$ structures, we subtracted a quadratic
background, Fig.1({\it b}).

The calculated $T(E)$ spectrum (minus a quadratic background) is also shown
in Fig.1({\it b}). A reasonable agreement with the experiment, especially
considering the neglect of umklapp, confirms our theoretical model of the
unoccupied states. The best averall agreement is achieved by shifting the
calculations by $\sim $3 eV to higher energies. Besides the computational
approximations, this can be attributed to excited-state self-energy
corrections $\Delta \Sigma $ to the DFT-GGA exchange-correlation. The
experiment suggests certain band dependence of $\Delta \Sigma $.

Guided by the calculated $T(E)$, we can trace the band structure origin of
the experimental $T(E)$ structures. The $T(E)$ maxima are seen to reflect
manifolds of so-called coupling bands (= dominant final states in PE) which
are characterized by large partial $T_{{\bf k}}$ currents \cite{CFS,CFC-Vg}.
The $T(E)$ minima reflect the gaps between such manifolds. Individual bands
are however not resolved in $T(E)$ because their energy separation is much
smaller than $%
\mathop{\rm Im}%
\Sigma $. Note that the coupling bands, apart from effective coupling to the
incident plane wave, must have a 3-dimensional (3D) character with
considerable $k_{\perp }$ dispersion to transmit current into the crystal;
they originate from the electronic orbitals oriented off-plane. The
2-dimensional (2D) bands, originating from the in-plane orbitals, are
ineffective in both VLEED and PE.

Finite electron lifetime manifests itself in broadening and damping of the $%
T(E)$ structures. We estimated the energy dependence of $%
\mathop{\rm Im}%
\Sigma $ based on its approximate parametric relation to the dielectric
function $\epsilon $ as $%
\mathop{\rm Im}%
\Sigma (E)=a+b\int_{0}^{E}%
\mathop{\rm Im}%
\left[ -1/\epsilon \left( {\bf q}=0,\omega \right) \right] d\omega $ with $%
\epsilon \left( {\bf q}=0,\omega \right) $ taken from the electron energy
loss data \cite{Norman99}. The parameters $a$ and $b$ were optimized to
bring the calculated $T(E)$ to the best agreement with the experiment in
broadening and damping of the spectral structures. The obtained $%
\mathop{\rm Im}%
\Sigma (E)$, which yields the calculated $T(E)$ in Fig.1 ({\it b}), is shown
in the insert in Fig.1 ({\it a}) superimposed on $\epsilon \left( {\bf q}%
=0,\omega \right) $. The increase of $%
\mathop{\rm Im}%
\Sigma $ results from two plasmon peaks in $\epsilon \left( {\bf q}=0,\omega
\right) $ associated with the CuO valence bands \cite{Norman99}. This
increase of $%
\mathop{\rm Im}%
\Sigma $ in the final states explains, for example, why the PE intensity
ratio of the CuO bilayer bonding and antibonding states varies with $h\nu $
above 30 eV much smoother in the experiment than in simulations employing
constant $%
\mathop{\rm Im}%
\Sigma =$1 eV \cite{Lee02} relevant closer to $h\nu $ of 20 eV.

Based on the experimental $%
\mathop{\rm Im}%
\Sigma (E)$ and calculated band dispersions, we found that the inelastic
electron scattering limits the mean free path $\lambda $ around 3\AA ,
decreasing with energy, in agreement with the electron energy loss data \cite%
{Norman99}. In the final-state band gaps $\lambda $ is further reduced by
elastic scattering off the crystal potential, resulting in a diffraction
structure in $\lambda (E)$, particularly strong where $%
\mathop{\rm Im}%
\Sigma $ is small. Qualitatively, it follows the structure in $T(E)$ where
the maxima reveal the 3D states enabling effective electron transport
(accurate evaluation of $\lambda $ involves calculations of damped Bloch
waves \cite{CFS,Krasovskii02}). Predictive tuning $\lambda $ based on the
experimental $T(E)$ can therefore be employed in the PE experiment to
resolve signals from different atomic layers.

Experimental dispersion of the $T(E)$ spectra with the incidence ${\bf K}%
_{\Vert }$ varying in the $\Gamma X$ azimuth (perpendicular to the IC
superstructure) is shown in Fig.2 (left). The shading shows the energy
intervals of the $T(E)$ maxima reflecting the coupling bands. This map gives
therefore the surface-projected ${\bf K}_{\Vert }$ dispersions of the final
states.

Our VLEED results give the first direct evidence of unoccupied quasiparticle
dispersions in Bi2212, whereas the inverse PE studies near $E_{F}$ (see,
e.g., \cite{Claessen89}) found only dispersionless states. Furthermore, the
highly structured non-parabolic dispersions illustrate a general fact that
at low excitation energies the final states, especially for complex
materials, dramatically deviate from a free-electron-like continuum due to
multiple scattering on the crystal potential. This property of the final
states is the essential origin of dramatic $h\nu $ dependences in PE.

The results for the $\Gamma Y$ azimuth (along the IC superstructure) are
shown in Fig.2 (right). By the fundamental lattice symmetry the $\Gamma Y$
azimuth is equivalent to $\Gamma X$, but this is complicated by the IC
superstructure. The $\Gamma Y$ dispersion image can then be viewed as the $%
\Gamma X$ dispersion with its IC\ umklapp replicas superimposed and
hybridized with the main dispersion. The observed image resembles therefore
a smeared $\Gamma X$ dispersion.

The dispersion in $\Gamma Y$ comes at first glance as a surprise: Competing
fundamental and IC potentials results in the collapse of strict periodicity
(and the entire band structure concept) and thereby the parallel momentum
conservation. As all states are now available for any incident ${\bf K}%
_{\Vert }$, one might expect that any dispersion in ${\bf K}_{\Vert }$ would
be lost. However, the VLEED spectra depend, apart from bare availability of
the states, on how their wavefunctions match the incident plane wave \cite%
{CFC-Vg}. Variation of the matching conditions with ${\bf K}_{\Vert }$
results in the observed dispersion. A similar appearance of competing
periodicities in PE was elucidated in \cite{Voit00}.

Our experimental $T(E)$ data yield thus the energy positions, ${\bf K}%
_{\Vert }$ dispersions and lifetimes of the PE final states. Apart from
direct implications in the PE experiment such as tuning $\lambda $ in energy
and ${\bf K}_{\Vert }$, this information can be used as a vital input in
computational simulations of the ME\ effects instrumental for resolving
details of the valence $E({\bf k})$ \cite{Lindroos02}. ME changes are
expected between the $T(E)$ maxima owing to a phase change in the
final-state wavefunction through the band gaps.

{\em Intensities of diffracted beams: hidden }$c${\em (2x2) periodicity and
final-state diffraction effects} -- A typical normal-incidence VLEED image
is shown in Fig.3 ({\it left}). Due to the IC superstructure the diffracted
intensity is arranged in lines along $\Gamma Y$ \cite{Claessen89,Lindberg88}%
, with non-vanishing intensity between the spots manifesting the fractal
character of diffraction in presence of two competing periodicities. The
intense spots, each with IC satellites, are seen to reflect, firstly, the
reciprocal lattice of the fundamental lattice at $(\pm 2\pi ,0)$ and $(0,\pm
2\pi )$. Surprisingly, there are other intense spots $(\pm \pi ,\pm \pi )$
(arrows), although weaker in $(-\pi ,\pi )$ and $(\pi ,-\pi )$. Their
symmetry reveals a hidden $c$(2x2) periodicity. This phenomenon is observed
in electron diffraction for the first time.

$I(V)$ data are shown in Fig.3 ({\it right}) as the VLEED intensity profile
along an IC diffraction line, as a function of energy and $q_{(-\pi ,\pi )}$
momentum transfer. All diffraction spots show a significant and
non-monotonous energy dependence beyond the kinematic picture. The $c$(2x2)
spots are intense only below $\sim $20 eV, and in the normal LEED energy
range they vanish.

To identify the origin of the $c$(2x2) periodicity, we first performed a
kinematic simulation using the IC superstructure parameters from \cite%
{Yamamoto90}. It has however yielded only the fundamental-lattice spots with
IC satellites. Beyond the kinematic picture, possible origins of the $c$%
(2x2) spots should be similar to the shadow FS with its $c$(2x2)
periodicity: either two inequivalent Cu sites per plane in the full
orthorhombic unit cell of Bi2212 \cite{Ding96}, or short-range AFM
correlations within the CuO planes \cite{Aebi94} manifesting themselves in
VLEED through the exchange scattering \cite{Palmberg68}. However, the
observed ${\bf K}_{\Vert }$ broadening of both $c$(2x2) spots and shadow FS
seems too small compared to the short AFM correlation lengths found by
neutron scattering. Finally, the $c$(2x2) spots can be peaks of multiple
scattering on the competing fundamental and IC\ potentials. Whichever the
origin, the diffraction process involves multiple scattering between the CuO
planes, and the outer SrO and BiO planes with their IC modulations \cite%
{Yamamoto90}. The involvement of the CuO planes is corroborated the $c$(2x2)
intensity reduction with energy which goes along with the decrease of $%
\lambda $.

Complementing the information on the final states obtained from $T(E)$, the
VLEED intensity distribution reflects their Fourier composition: Being the
time-reversed LEED state, the final state can be expanded over the surface
reciprocal vectors $\Phi \left( {\bf r}\right) =\sum_{{\bf g}}\phi _{{\bf g}%
}\left( r_{\bot }\right) e^{i\left( {\bf k}_{\Vert }+{\bf g}\right) {\bf r}%
_{\Vert }}$, where the Fourier coefficients $\phi _{{\bf g}}\left( r_{\bot
}\right) $ correspond, in vacuum, to the incident and diffracted plane waves
and, in the crystal, to the Bloch wavefield. The VLEED intensities $I_{{\bf g%
}}$ are the vacuum asymptotics of $\left| \phi _{{\bf g}}\left( r_{\bot
}\right) \right| ^{2}$. As $\phi _{{\bf g}}\left( r_{\bot }\right) $ run
across the surface continuously in amplitude and derivative, $I_{{\bf g}}$
reflect, approximately, the Fourier composition of the final state in the
crystal.

In the PE experiment this information immediately characterizes the
final-state diffraction effects. For example, in the FS mapping experiment a 
${\bf g}$-diffraction replica of the main FS occurs whenever upon variation
of ${\bf k}_{\Vert }$ the corresponing ${\bf k}_{\Vert }+{\bf g}$ Fourier
component of the final state matches, in the parallel momentum, any Fourier
component of the initial state, giving rise to non-zero overlap of the
final- and initial-state wavefunctions. The PE signal is then proportional,
apart from phases of the Fourier components and selection rules entering the
ME \cite{Lindroos99,Lindroos02,Borisenko00}, to the final-state $\left| \phi
_{{\bf g}}\left( r_{\bot }\right) \right| ^{2}$ and thus to its $I_{{\bf g}}$
asymptotics. Therefore, the PE signal of certain ${\bf g}$-diffraction
replica follows, roughly, the ${\bf g}$-diffracted intensity observed in the
VLEED image. The $I(V)$ data can then be used to tune the final-state energy
to enhance or suppress certain diffraction replicas in the FS mapping
experiment. Similar ideas can be exploited to single certain bands out of
the valence band multitude based on their dominant Fourier component. Apart
from Bi2212, such an approach can be applied to detail electronic structures
of other complex materials, for example, perovskites.

Our VLEED data on the final-state diffraction effects leads to significant
revisions in the current interpretation of some PE data on Bi2212. For
example, the shadow FS with its $c$(2x2) periodicity is commonly considered
as an intrinsic effect pertinent to the initial state. To check against the
extrinsic effects due to the final-state diffraction, we have measured VLEED
images with incidence ${\bf K}_{\Vert }$ set near the $M$ point where a
large shadow FS signal is found in the PE experiment. The $c$(2x2) spots
were found intense below $\sim $25 eV and gradually vanishing at higher
energies. This indicates that at low $h\nu $, commonly used in the PE
experiment, the shadow FS intensity has a large extrinsic contribution. This
is corroborated by the absence of a hybridization gap between the main and
shadow FS \cite{Borisenko00}. The intrinsic shadow FS signal should be
measured with higher $h\nu $ above 25 eV.

Finally, it should be noted that the final-state energies in PE, strictly
speaking, differ from those observed in VLEED by the electron-hole
interaction (excitonic) energy $E_{e-h}$. This figure, characterizing
deviations from the sudden approximation, can be determined by comparing
characteristic points in the PE intensity behavior with the predictions of
VLEED. For typical layered materials such an analysis has however yielded
negligible $E_{e-h}$ values, at least on the scale of 100 meV \cite{TMDCs}.

{\em Conclusion} -- Using VLEED with a support of full-potential band
calculations, we have achieved first direct information on the energy
positions, ${\bf K}_{\Vert }$ dispersions, lifetimes and Fourier composition
of the PE final states in Bi2212 at the excitation energies up to 45 eV
above $E_{F}$. Implications of this information for the PE experiment extend
from tuning $\lambda $ to predictive use of the ME effects to focus on
specific valence states or FS\ replicas. In particular, intense final-state $%
c$(2x2) diffraction identifies significant extrinsic contribution to the
shadow FS.

We thank M. Golden, A. Zakharov and E.E. Krasovskii for promoting
discussions. The work was supported by Deutsche Forschungsgemeinschaft
(grant Cl 124/5-1).

\begin{figure}[tbp]
\caption{({\it a}) Experimental $T(E)$ spectrum; ({\it b}) Structure in the
experimental and calculated $T(E)$ enhanced by subtracting a quadratic
background. The calculation is shifted by +3 eV to mimic $\Delta \Sigma $.
The inset shows the experimental $%
\mathop{\rm Im}%
\Sigma $ (with the uncertainty bars) compared with $%
\mathop{\rm Im}%
\left[ -1/\protect\varepsilon \right] $; ({\it c}) Calculated $E({\bf k})$
with the partial currents $T_{{\bf k}}$ shown by grayscale. The $T(E)$
maxima reflect manifolds of the final bands coupling to vacuum.}
\end{figure}

\begin{figure}[tbp]
\caption{Experimental ${\bf K}_{\Vert }$ dispersion of the $T(E)$ spectra.
The shading shows the energy regions of the $T(E)$ maxima, identified by $%
d^{2}T/dE^{2}>0$. The values of $d^{2}T/dE^{2}$ in these regions are
represented in a logarithmic grayscale (black = maximal value). This map is
the surface-projected dispersion of the PE final bands.}
\end{figure}

\begin{figure}[tbp]
\caption{({\it left}) Normal-incidence VLEED image at an energy of 16.4 eV
above $E_{F}$. The specular beam is in the center obscured by the electron
gun. The diffracted intensity maxima reflect the fundamental lattice and a
hidden $c$(2x2) periodicity (arrows); ({\it right}) $I(V)$ data for the
indicated IC diffraction line shown in a linear grayscale (white = maximal
intensity). The coordinates on the LEED screen, somewhat distorted by the
retarding field, are rendered into the $q_{(\protect\pi ,-\protect\pi )}$
values.}
\end{figure}

\end{document}